\documentclass[onecolumn]{revtex4}
\usepackage{amsmath}
\usepackage{epsfig}
\makeatletter
\def\btt#1{\texttt{\@backslashchar#1}}%
\DeclareRobustCommand\bblash{\btt{\@backslashchar}}%
\makeatother

\begin{document}

\title{QUASINORMAL MODES AND LATE-TIME TAILS OF CANONICAL
ACOUSTIC BLACK HOLES}

\author{Ping Xi}
\author{Xin-Zhou Li}\email{kychz@shnu.edu.cn}
\affiliation{Shanghai United Center for Astrophysics(SUCA),\\
 Shanghai Normal University, 100 Guilin Road, Shanghai 200234,China\\
 and\\
School of Science,\\
East China University for Science and Technology, 130 Meilong Road
Shanghai, 200237, China
}%

\date{\today}

\begin{abstract}
\section*{abstract}
In this paper, we investigate the evolution of classical wave
propagation in the canonical acoustic black hole by numerical method
and discuss the details of tail phenomenon. The oscillating
frequency and damping time scale both increase with the angular
momentum $l$. For the lower $l$, numerical results show the lowest
WKB approximation gives the most reliable result. We also find that
time scale of the interim region from ringing to tail is not
affected obviously by changing $l$.
\end{abstract}


\maketitle
\section{Introduction}
 Some properties of black holes can be investigated using acoustic
analogues in the laboratory through the propagation of sound wave.
Hawking radiation is a remarkable prediction and is almost
universally believed to be one of the most important in black hole's
physics. However, the Hawking temperature of astrophysical black
holes is much smaller than the temperature of the cosmic microwave
background so that one cannot acquire any conclusive evidence of the
existence of Hawking radiation. About twenty-five years ago, Unruh
proposed a method that certain aspects of astrophysical black hole
are mapped into problems in the theory of supersonic acoustic flows
\cite{EBHE}. Even though the Hawking temperatures associated to
acoustic analogues are not high enough to be detectable up to now,
the situation is likely to change in the near future
\cite{acoustic}. A profound understanding of the classical physics
of acoustic black hole is indispensable for the detection of Hawking
radiation. Berti, Cardoso and Lemos \cite{Berti} investigated wave
propagation in the ``draining bathtub'' model and the ``canonical''
(1+3)-dimensional acoustic black hole \cite{Visser}. Especially,
using the Wentzel-Kramers-Brillouin (WKB) method, they calculated
the quasinormal modes (QNMs). Many physicists believe that the
figure of QNMs is a significant fingerprint indirectly identifying
the existence of a black hole. The QNMs of black holes in the
framework of general relativity \cite{Cardoso,Giammatteo} and string
theory \cite{Li,Xi} has been studied widely.

Approximately, there are three stages in the evolution of the
perturbations of an acoustic black hole \cite{Berti}. First stage is
the rapid response at very early time, on which the initial
conditions have a great effect. Second stage is quasinormal ringing
phase, which characteristic oscillation frequencies and damping
times depend strongly on the acoustic analogue QNMs. The QNMs are
determined completely by the parameters of system, therefore they
would carry significant information about the background curvature
of the intervening spacetime. Finally, there is a tail stage, which
decays approximately as a power in time owing to backscattering off
the spacetime curvature. In Ref. 3, the authors have used three WKB
computational schemes, i.e. the lowest approximation \cite{Will},
3rd order improvements \cite{Iyer,Seidel} and 6th order corrections
\cite{Konoplya}. For the canonical acoustic black hole, the results
show that $l=1$ QN frequencies seem to be the problem, in which the
mode suffers a large variation as one goes from the lower
approximation to the higher approximation. This means that the WKB
approach is more dependable for higher $l$, which was first
discovered in the early work \cite{Will,Iyer,Seidel}. Therefore, the
numerical calculation is necessary for the lower $l$ QN
frequencies.\\
\indent In this paper, we investigate in detail the relations
between QNMs of canonical acoustic black hole and the angular
momentum $l$ by the numerical calculation in null coordinates. Some
results attained by this way are supported by the analytic results
and WKB results. Most of importance, we confirm that the lowest WKB
approximation gives the most reliable results for $l=1$ case.
Furthermore, we show a picture of classical wave propagation
including the interim region from the quasinormal ringing to tail
stage.

\section{FORMALISM AND BASIC EQUATIONS}

Assume the fluid to be incompressible and spherically symmetric,
then since background density $\rho $ is position independent the
continuity equation implies the velocity is in proportion to $r^{ -
2}$. The background pressure $p$ and speed of sound $c$ are also
position independent because of the barotropic assumption.
Therefore, one can define a normalization constant $r_{0} \equiv
\left( {vr^{2} / c} \right)^{{\frac{{1}}{{2}}}}$. The canonical
acoustic metric describing the propagation of sound waves in this
incompressible and spherically symmetric (1+3)-dimensional fluid
flow \cite{Visser} is:

\begin{equation}
\label{eq1} ds^{2} = - c^{2}\left( {1 - {\frac{{r_{0}
^{4}}}{{r^{4}}}}} \right)dt^{2} + \left( {1 - {\frac{{r_{0}
^{4}}}{{r^{4}}}}} \right)^{ - 1}dr^{2} + r^{2}(d\theta ^{2} + \sin
^{2}\theta d\phi ^{2})
\end{equation}
The metric (\ref{eq1}) is distinct from any of the geometries
typically considered in general relativity. Unruh \cite{EBHE} first
suggested that the propagation of a sound wave is described by the
Klein-Gordon equation $\nabla _{\mu}  \nabla ^{\mu} \Psi = 0$ for a
massless scalar field $\Psi $ in a Lorentzian acoustic geometry,
which take metric (\ref{eq1}) at present. We can separate variables
by setting

\begin{equation}
\label{eq2} \Psi (t,r,\theta ,\phi ) = {\frac{{1}}{{r}}}\Phi (r_{
\ast} ,t)Y_{lm} (\theta ,\phi )
\end{equation}
where $Y_{lm}(\theta,\phi)$ are the usual spherical harmonics and
the tortoise coordinate $r_{\ast}$ is defined by

\begin{equation}
\label{eq3}
r_{\ast}={\int{(1-{\frac{{r_{0}^{4}}}{{r^{4}}}})^{-1}dr}}
\end{equation}

\noindent where we have chosen unit $c=1$. The evolution equation of
$\Phi(r_{\ast})$ is

\begin{equation}
\label{eq4} {-\frac{{\partial ^{2}\Phi}}{{\partial
t^{2}}}+\frac{{\partial ^{2}\Phi}}{{\partial r_{\ast}^2}}}=V\Phi
\end{equation}

\noindent where the effective potential

\begin{equation}
\label{eq5}
V(r_{\ast})=(1-{\frac{{r_{0}^{4}}}{{r^{4}}}})[{\frac{{l(l+1)}}{{r^{2}}}}+{\frac{{4r_{0}^4}}{{r^{6}}}}]
\end{equation}

We introduce the null coordinates $u=t-r_{\ast}$ and $v=t+r_{\ast}$,
Eq. (\ref{eq5}) can be reduced to

\begin{equation}
\label{eq6} -4{\frac{{\partial ^{2}\Phi}}{{\partial u \partial
v}}}=V(r_{\ast})\Phi
\end{equation}

\noindent Eq. (\ref{eq6}) can be numerically integrated by the
ordinary finite element method. Using the Taylor expansion, we have

\begin{equation}
\Phi_{N}=\Phi_{E}+\Phi_{W}-\Phi_{S} -\delta u \delta
v(\frac{v_{N}+v_{W}-u_{N}-u_{E}}{4})\frac{\Phi_{W}+\Phi_{E}}{8}V(r_{\ast})
+O(\Delta^{4})
\end{equation}

\noindent where $N$, $W$, $E$ and $S$ are the points of a unit grid
on the $u-v$ plane which correspond to ($u+\Delta$,$v+\Delta$
),($u+\Delta$,$v$),($u$,$v+\Delta$) and ($u$, $v$), and $\Delta$ is
the step length of the change of $u$ or $v$, i.e., $\Delta=\delta
u=\delta v$ \cite{Li,Xi}. Because the quasinormal ringing stage and
the late time stage are both insensitive to the initial conditions,
we begin with a Gaussian pulse of width $\sigma$ centred on $v_{c}$
when $u=u_{0}$ and set the field $\Phi$ is zero on $v=v_{0}$,

\begin{eqnarray}
\Phi(u=u_{0},v)&=&exp[-{\frac{{(v-v_{c})^{2}}}{{2 \sigma^{2}}}}] \nonumber \\
\Phi(u,v=v_{0})&=&0
\end{eqnarray}
Next, the point in the $u-v$ plane can be calculated by using Eq.
(\ref{eq6}), successively. Finally, the values of $\Phi(u_{max},v)$
are extracted after the integration is completed where $u_{max}$
represents the maximum of $u$. Taking sufficiently large $u_{max}$
for the various $v$-value, we obtain a good approximation for the
wavefunction of canonical acoustic black hole.

\section{NUMERICAL RESULTS}
Our numerical results, which are all consistent with the analytic
results and WKB results in Ref. 3, are shown in Figs. 1-6. As a
reminder, the oscillating period, damping time scale and late time
tail are shown in these figures. Here, the parameter $r_{0}$ is set
to uninty. The dependence of quasinormal modes on $r_{0}$ is
trivial. On the one hand, the canonical acoustic metric coordinates
can be rescaled to set $r_{0}=1$. On the other hand, the results
must depend linearly on $r_{0}$ since it is the only dimensional
quantity in the problem. In Fig. 1, we show the relations between
the wavefunction and the angular momentum $l$. Our numerical result
is consistent with Ref. 3. That means the oscillating period and the
damping frequency both decrease when the index $l$ increases.
Furthermore, we confirm that the lowest WKB approximation gives the
most reliable results for $l=1$ case. To further corroborate this
conclusion, we list QN frequencies for $l = 1, 2, 3, 4$ in Table
1.\\
\begin{table}[ph]
{QN frequencies of the canonical acoustic black hole for $l = 1,
2, 3, 4$ are listed.}\\
\begin{tabular}{@{}cccc@{}}\toprule $l$ &
$Re(\omega)$ &$Im(\omega)$
 \\ \colrule
1\hphantom{0} & \hphantom{0}1.463 & \hphantom{0}0.666  \\
2\hphantom{0} & \hphantom{0}1.619 & \hphantom{0}0.653  \\
3\hphantom{0} & \hphantom{0}1.642 & \hphantom{0}0.625  \\
4\hphantom{0} & \hphantom{0}1.758 & \hphantom{0}0.620  \\ \botrule
\end{tabular}
\end{table}

In Figs. 2-6, we choose $l=2$, and consider in detail the picture of
classical wave propagation in the canonical three stages, the second
and final stages as illustrated in Fig. 2. The prompt contribution
is the evident counterpart of light cone propagation in the $V=0$
case, which strongly depends on the initial conditions, therefore it
is left out in Fig. 2. At intermediate $v$ values the wavefunction
is dominated by an exponential decay, whose oscillation frequency
and damping time are described by its QNMs. At the late-time (large
$v$ value) the propagating wave leaves a power-law tail which is
magnified in Fig. 3. By numerical calculation, we attain the
expression of power-law falloff, $\Phi \approx 7.36\times
10^{-32}t^{-10}$, which is consistent with the analytic result in
Ref. 3. Especially, the interim region from ringing stage to tail
stage is corresponding to rectangular region A in Fig. 2. This
interim region is replotted in Fig. 4, where the $v$-coordinate is
magnified about $10^{3}$ times. The time interval from region B to
region C is so short that the numerical results between region B and
region C seem unfaithful. Therefore, we do not discuss physical
implications about this region, attentively. The rectangular regions
B and C of Fig. 4 are magnified in Fig. 5 and Fig. 6, respectively.
In region B, the oscillation frequency dramatically changes and
tends to zero. Likewise, in region C, the damping time scale also
has a drastic change, which becomes infinity. These figures tell us
how perturbation in vicinities of this black hole die out as a
late-time tail. It is easy to find that time scale of interim region
from ringing to tail is not affected obviously
by changing the angular momentum $l$.\\
\begin{figure}[pb]
\begin{center}
\epsfig{file=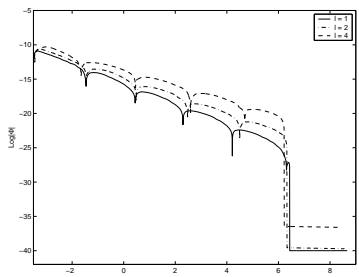,width=4.7cm} \caption{For the canonical
acoustic black hole with $r_{0}=1$, the wavefunctions are shown via
different the angular momentum $l$. The logarithm is to base 10.}
\end{center}
\end{figure}
\begin{figure}[pb]
\begin{center}
\epsfig{file=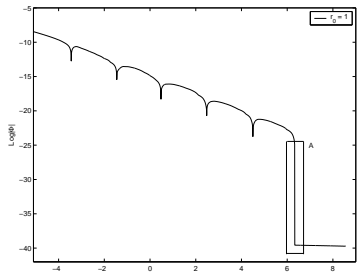,width=4.7cm} \caption{For the canonical
acoustic black hole with $r_{0}=1$ and $l=2$, the rectangular region
A describes as interim region from ringing to tail. The logarithm is
to base 10.}
\end{center}
\end{figure}
\begin{figure}[pb]
\begin{center}
\epsfig{file=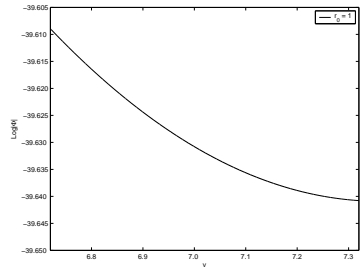,width=4.7cm} \caption{The late-time tail in
Fig. 2 is magnified. The logarithm is to base 10.}
\end{center}
\end{figure}
\begin{figure}[pb]
\begin{center}
\epsfig{file=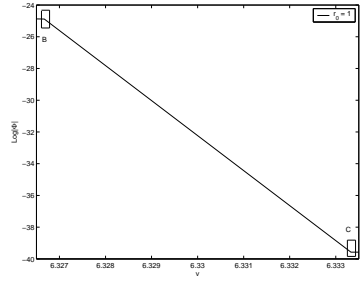,width=4.7cm} \caption{For the canonical
acoustic black hole with $r_{0}=1$ and $l=2$, the interim region
from ringing to tail is shown.}
\end{center}
\end{figure}
\begin{figure}[pb]
\begin{center}
\epsfig{file=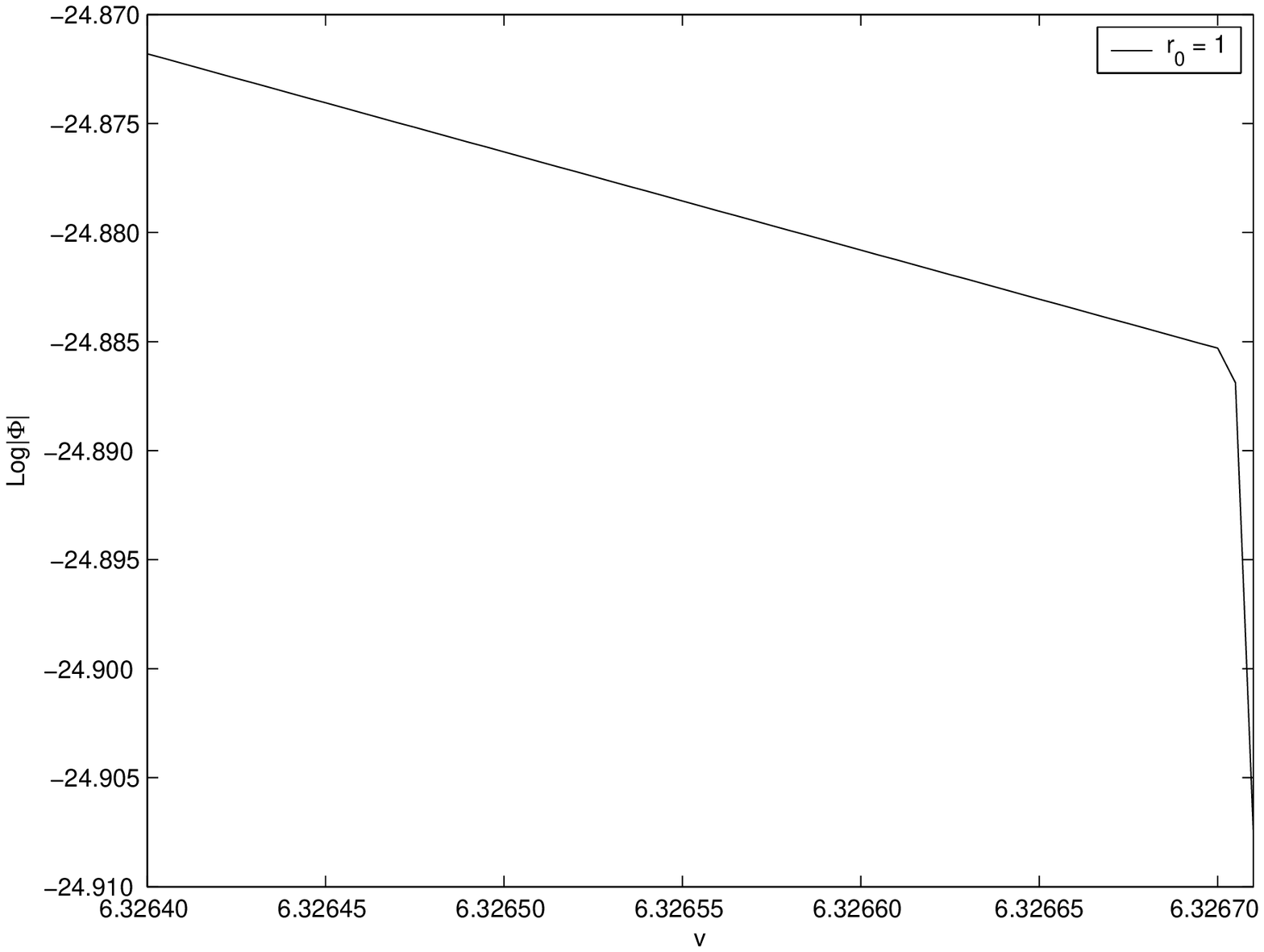,width=4.7cm} \caption{The rectangular
region B in Fig. 4 is magnified. The logarithm is to base 10.}
\end{center}
\end{figure}
\begin{figure}[pb]
\begin{center}
\epsfig{file=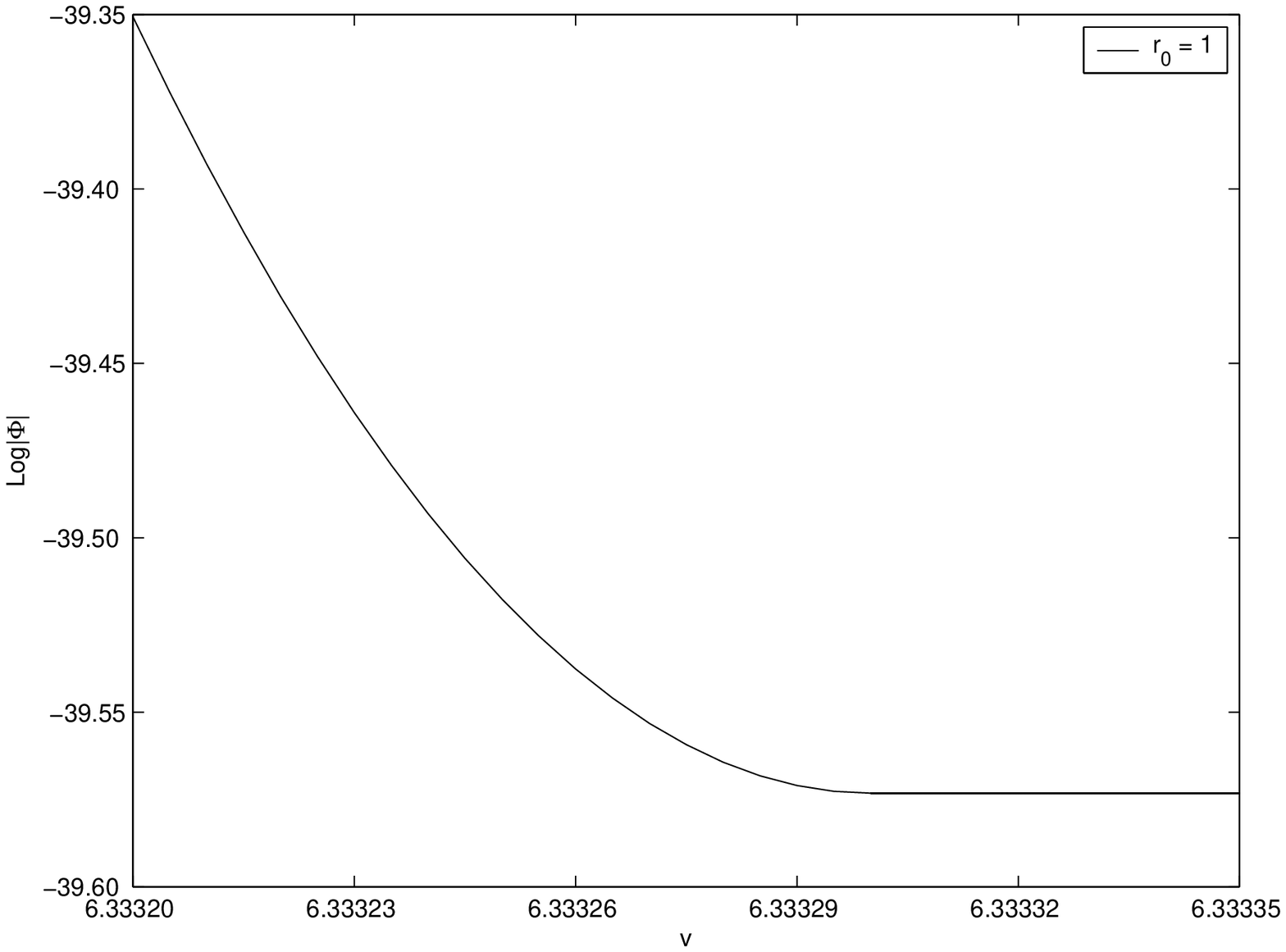,width=4.7cm} \caption{The rectangular
region C in Fig. 4 is magnified. The logarithm is to base 10.}
\end{center}
\end{figure}

\section{CONCLUSIONS}

In this work we considered numerically the evolution of classical
wave propagation in the canonical acoustic black hole and discussed
the details of tail phenomenon. We summarize main results as follows:\\
\indent(i) For $l\geq 2$, the numerical results are consistent with
the first, third and sixth order WKB method. For the lower $l$,
numerical results show the first WKB approximation gives the most
reliable result because of the basic WKB assumption (the ratio of
the derivatives of the potential to the potential itself should be
small) is broken. \\
\indent(ii) From a physical viewpoint, the most reasonable
explanation for the production of late-time tails is the
backscattering of waves off a spacetime curvature at asymptotically
far regions. Our numerical results show that late-time tail is
consistent with Ref. 13, and time scale of the interim region from
ringing to tail is not affected obviously by changing the angular
momentum $l$.\\
\indent(iii) The oscillating frequency and damping time scale both
increase with the angular momentum $l$. In the limit of large $l$,
the real part of fundamental QN frequency increase linearly and
imaginary part tend to a constant with the angular momentum $l$.

\section*{Acknowledgments}

This work was partially supported by the National Nature Science
Foundation of China under Grant No. 10473007.

\end{document}